\pdfoutput=1

\documentclass[aps,prl,preprint,groupedaddress]{revtex4}
\usepackage{graphicx}
\usepackage{subfigure}

\begin{document}


\title{Oxide layer boron leads to reduced symmetry spin filtering magnetic tunnel junctions}


\author{Derek A. Stewart}
\email[]{stewart@cnf.cornell.edu}
\affiliation{Cornell Nanoscale Science and Technology Facility, Ithaca, NY 14853}


\date{\today}

\begin{abstract}
Experimental studies of FeCoB/MgO/FeCoB tunnel junctions indicate that boron diffuses into MgO during rf-sputtering and forms polycrystalline Mg-B-O regions.  These tunnel junctions provide high tunneling magnetoresistance values and low RA products.  However the crystal structure of the Mg-B-O region remains unknown.  Using density functional techniques, I examine three potential Mg(B) oxides including Mg$_{2}$B$_{2}$O$_{5}$ (monoclinic and triclinic) and the orthorhombic mineral Kotoite (Mg$_3$B$_2$O$_6$).  Kotoite is the best candidate for formation in magnetic tunnel junctions.  The (100) surface of Kotoite has a good lattice match with (001) MgO and could template neighboring FeCo into bcc layers during annealing.  Complex band structure analysis of Kotoite shows that the C$_{2v}$ $\tilde{\Delta}_1$ band has a much smaller imaginary $k$ component than the C$_{2v}$ $\tilde{\Delta}_4$ band.  Based on symmetry analysis, the majority spin $\Delta_1$ band in FeCo should couple well with the Kotoite $\tilde{\Delta}_1$ band, while the minority FeCo $\Delta_5$ will couple partially with the $\tilde{\Delta}_4$ band.  Kotoite provides a new route to high tunneling magnetoresistance based on spin filtering by a lower symmetry oxide region.
\end{abstract}


\maketitle

Magnetic tunnel junctions (MTJs) based on sputtered amorphous CoFeB leads and MgO tunnel barriers exhibit high tunneling magnetoresistance (TMR) at room temperature (604\%)\cite{ikeda_600_tmr_apl_2008}.  High TMR values coupled with low RA products are essential for memory storage devices and CoFeB/MgO/CoFeB MTJs with low RA products have recently been demonstrated\cite{isogami_2008_high_tmr_low_ra}.  
Boron ensures that deposited CoFe leads are amorphous, reducing interface roughness\cite{dyajaprawira_2005} and allowing MTJs to grow on synthetic antiferromagnets\cite{yuasa_mtj_review_2007}.  Upon annealing, MgO acts as a template for the growth of crystalline bcc FeCo layers on either side of the oxide\cite{yuasa_2005_mgo_template} and provides the epitaxial interface required for spin filtering in the oxide layer\cite{butler_prb_2001}.

While CoFeB leads have been very successful in improving MTJ properties, the exact role played by boron and where it resides after sputtering and subsequent annealing has remained a mystery.  Originally it was assumed that all boron diffused away from the MgO layer during annealing.  This would leave isolated FeCo layers at the MgO interface, resulting in high TMR values.  However, recent work indicates that the movement of boron can depend strongly on deposition and annealing conditions.  A theoretical study found that a single layer of boron at the CoFe/MgO interface was energetically favorable and could significantly reduce TMR\cite{burton_2006}.  However, this work did not examine the possibility of boron diffusion and incorporation into MgO.  Recent experimental studies provide evidence that boron oxide does form during sputtering deposition\cite{bae_2006,read_2007,petford_long_apl_2008} and that a significant amount of boron can diffuse into MgO during annealing and enhance TMR for thin oxide tunnel junctions\cite{read_mgbo_barriers_apl_2009}.  Mg and O K-edge EELS data from these studies indicated that Mg and O coordination was different from that of bulk MgO, indicating the presence of Mg-B oxides.  Boron K-edges in the oxide region show that boron is oxidized with a BO$_3$ configuration that is present after deposition and remains even after annealing\cite{cha_eels_apl_2007}.  

Experimental studies of MTJs with Mg-B-O oxide regions show that boron incorporation into the oxide region (1) preserves a high TMR value upon annealing, (2) provides MTJs with the low RA product needed for industrial applications, and (3) acts to template the neighboring FeCo leads to grow in bcc crystalline layers.  However, a clear understanding of the specific Mg-B-O crystal structure in this region is crucial for understanding the underlying physical properties of this system and how they may be optimized.  I examine how the incorporation of boron affects the local density of states and use complex band structure analysis to address whether symmetry based spin filtering is still possible in these Mg-B-O oxides.

\begin{figure}
\begin{center}
   \mbox{
   \subfigure[ ]{
   \label{fig1a}
   \includegraphics[angle=0,width=6cm]{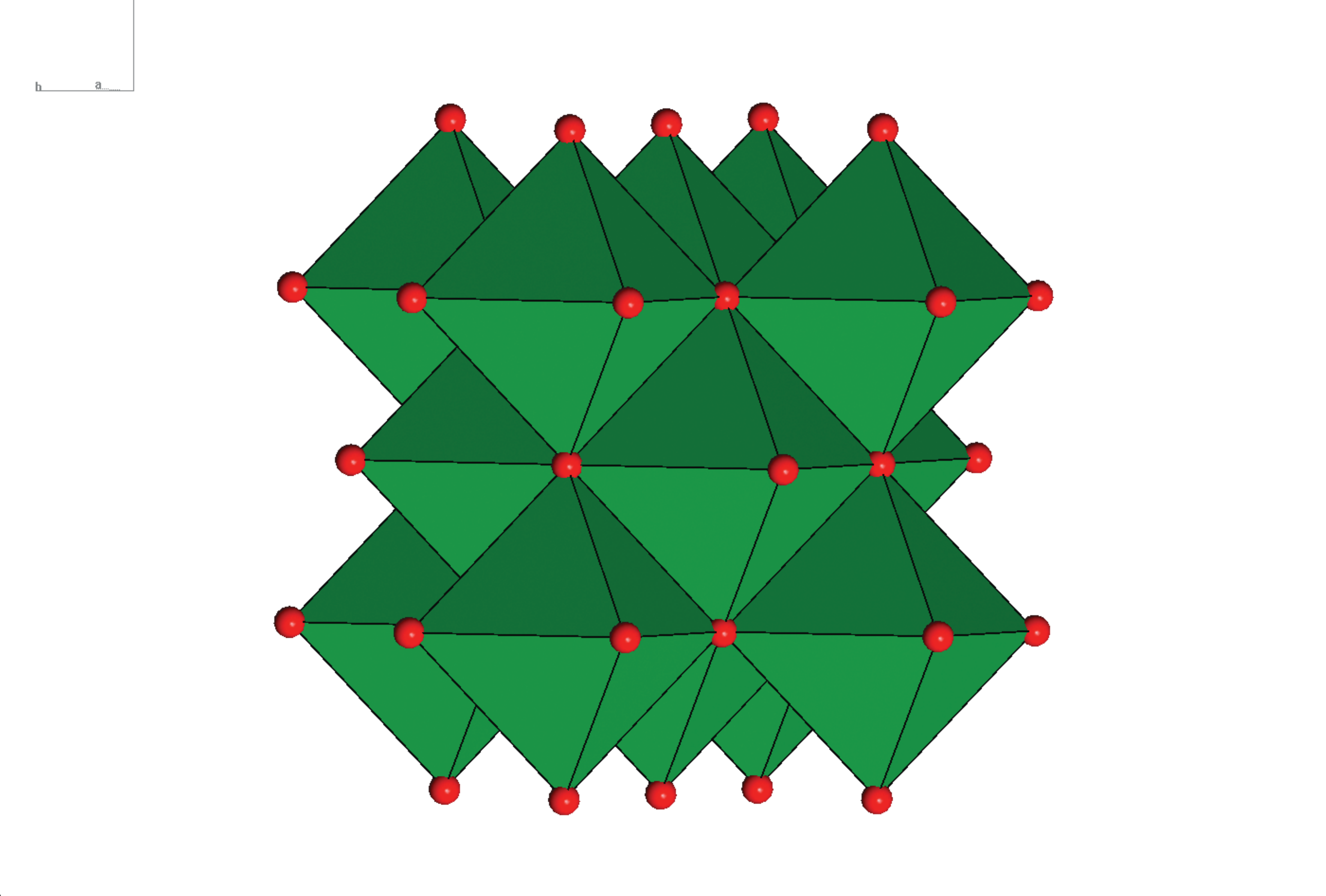} 
   } 
   \subfigure[ ]{
   \label{fig1b}
   \includegraphics[angle=0,width=6cm]{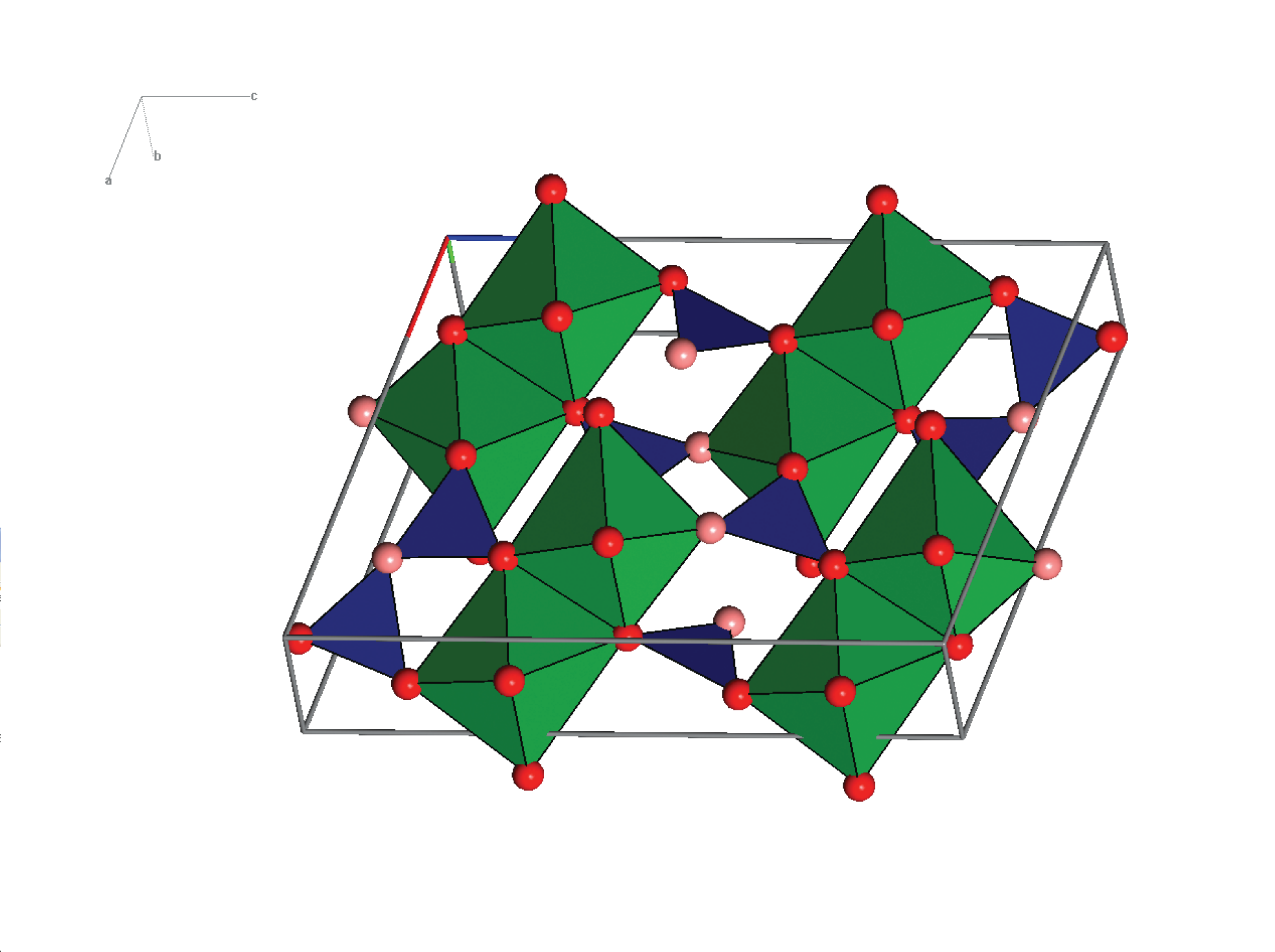}
   }
   }
   \mbox{
   \subfigure[ ]{
   \label{fig1c}
   \includegraphics[angle=0,width=6cm]{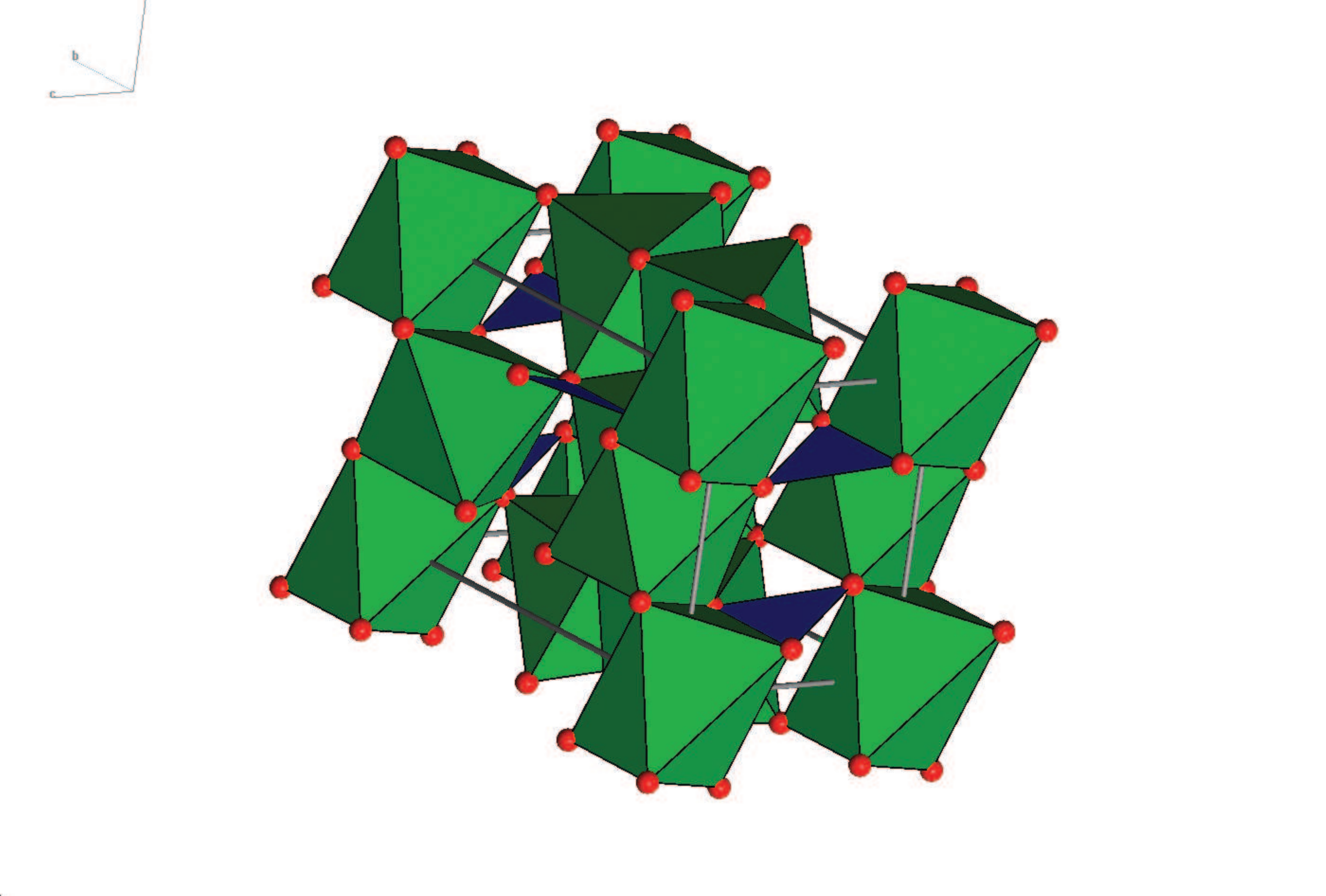}
   }
   \subfigure[ ]{
   \label{fig1d}
   \includegraphics[angle=0,width=6cm]{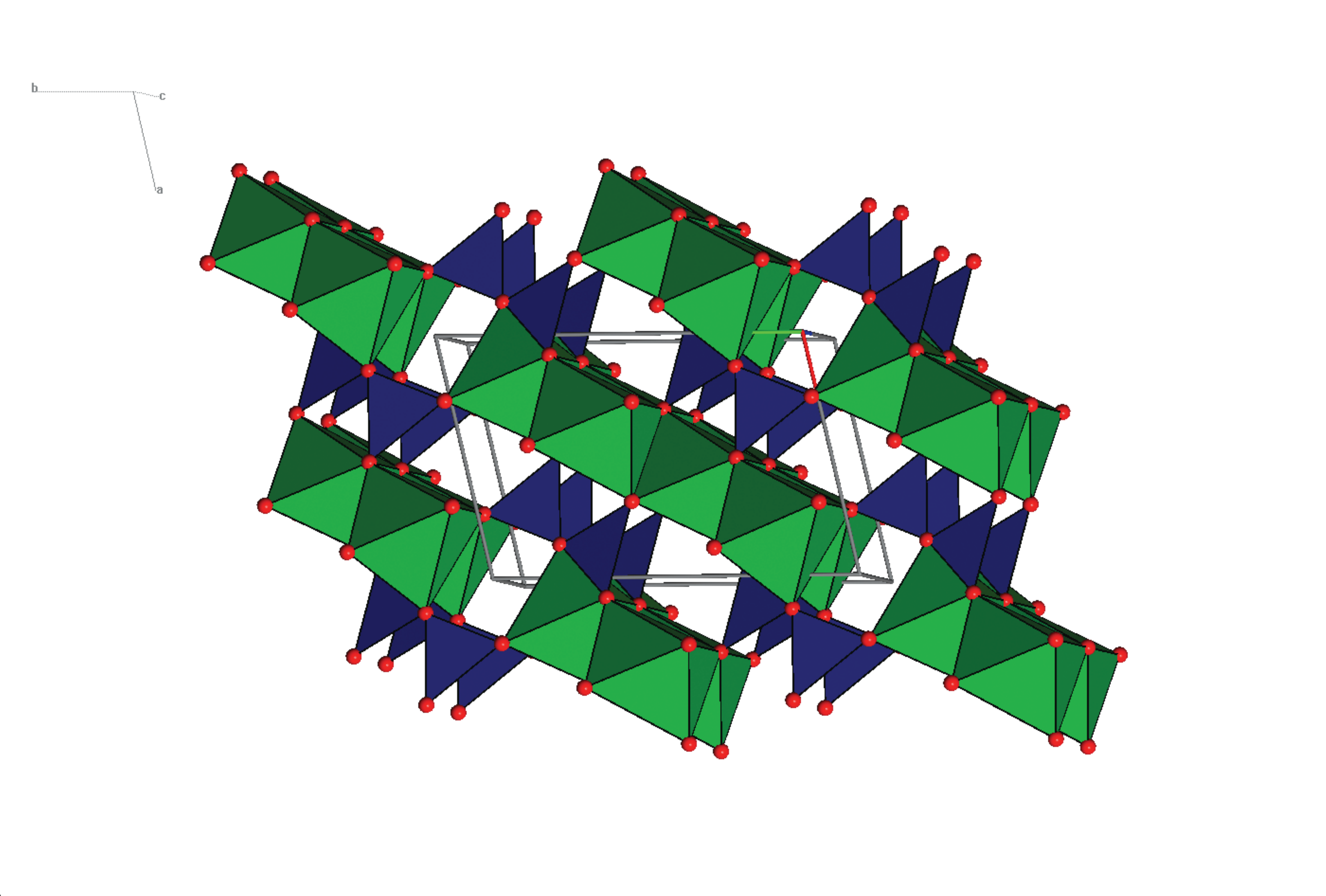}
   }
   }
\caption{(color online) The crystal structures for (a) MgO, (b) Mg$_2$B$_2$O$_5$ (monoclinic), (c) Kotoite Mg$_3$B$_2$O$_6$, and (d) Mg$_2$B$_2$O$_5$ (triclinic).  Oxygen atoms are given in red and magnesium atoms are located at the center of the green octahedra.  Boron atoms are located at the center of the blue triangles.}  
    \label{all_crystal_structures}
	\end{center}
\end{figure}

In this Letter, I investigate potential boron based oxide formation within tunnel junctions.  I consider three candidate Mg-B oxides including magnesium borate (Mg$_{2}$B$_{2}$O$_5$) in its monoclinic\cite{mono_mg2b2o5_acta_cryst} (space group P2$_1$/c) and triclinic\cite{triclinic_acta_mater} (space group P$\bar{1}$) forms as well as the mineral Kotoite (Mg$_3$B$_2$O$_6$) (space group P$_{nmn}$)\cite{kotoite_crystal} (see Fig. ~\ref{all_crystal_structures}).  In each crystal, the Mg atoms are surrounded by an octahedron of oxygen atoms, shown in green in Fig.~\ref{all_crystal_structures}.  Boron atoms are always surrounded by three oxygens, BO$_3$, in a triangular configuration shown as blue triangles.  The main difference between Mg$_2$B$_{2}$O$_5$ and Mg$_2$B$_{3}$O$_6$ comes from the coordination of BO$_3$ triangles.  In Kotoite (Mg$_3$B$_{2}$O$_6$), single BO$_3$ triangles link chains of oxygen octahedra, while in Mg$_2$B$_{2}$O$_5$ crystal structures, double BO$_3$ triangles link together oxygen octahedra\cite{boron_class}.  While the monoclinic form of Mg$_{2}$B$_{2}$O$_5$ is found in nature as the rare mineral Suanite\cite{mono_mg2b2o5_acta_cryst}, synthetically produced magnesium borate is used in industry as a catalyst for hydrocarbon conversion\cite{qasrawi_2005} and for friction reduction\cite{mgbo_wear_2002}.  Monoclinic and triclinic Mg$_2$B$_{2}$O$_5$ nanowires have also been grown\cite{mg2b2o5_nanowires_nl,mg2b2o5_nanowires_chem_mater}.      

I examined the impact of boron on the 
density of states (DOS) in the oxide region.  Density functional calculations were performed using both a pseudopotential-plane wave approach (as implemented in Quantum Espresso\cite{pwscf}) and a full potential augmented spherical wave approach\cite{asw_eyert}.  Since very little work has been done on the electronic structure of these systems, using multiple \textit{ab-initio} techniques provides an additional accuracy check.  All calculations were done within the local density approximation.  A 100 Ryd cutoff energy was used for plane wave calculations.  Monkhorst-Pack k-point grids of 12x12x12, 8x16x8, 10x6x12, and 8x4x12 were used for MgO, Mg$_2$B$_2$O$_5$ (monoclinic), Mg$_2$B$_2$O$_5$ (triclinic), and Kotoite, respectively.  Mg and B ions were described using von Barth Car pseudopotentials\cite{vonbarth_car}, while an ultrasoft pseudopotential was used for O atoms\cite{ultrasoft_vanderbilt}.  Exchange and correlation interactions were represented by the Perdew and Zunger parameterization\cite{perdew_zunger_ex}.  Lattice constants and atomic positions were relaxed based on Hellman-Feynman forces.  For augmented spherical wave calculations, empty spheres were added to insure a proper description of open regions in the crystal structure. 



\begin{figure}
\begin{center}
\centering
\includegraphics[angle=0,width=12.00cm]{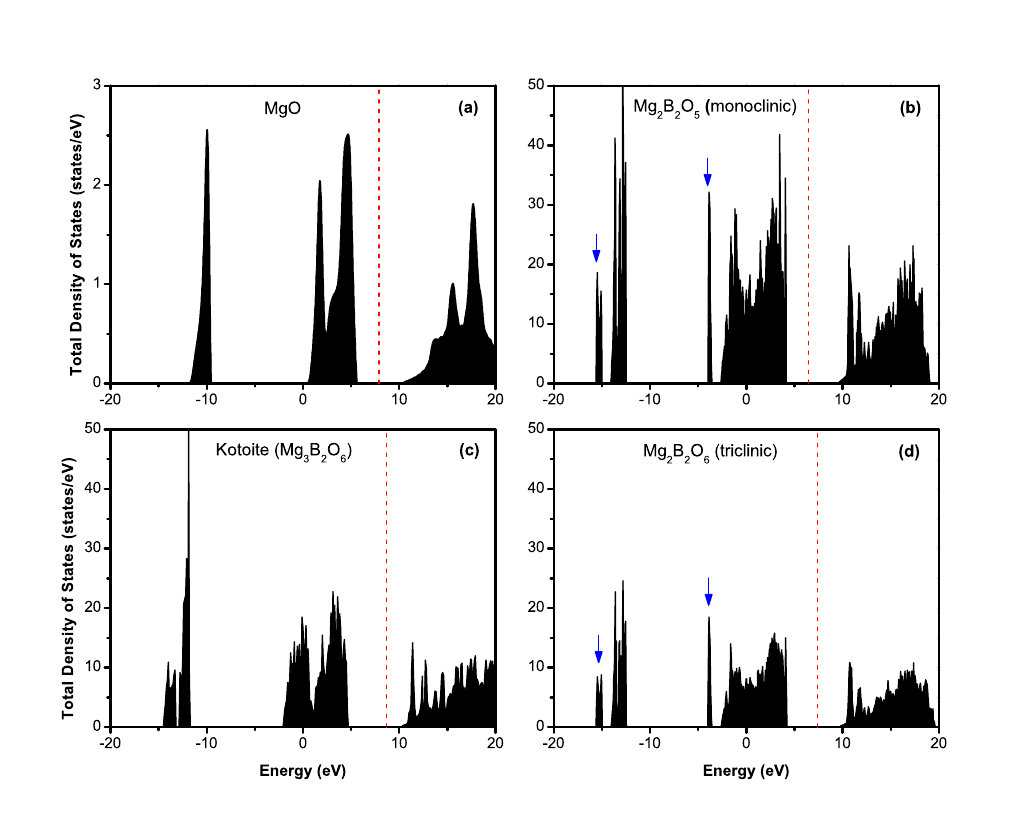}
\caption{(color online) The total density of states is shown for (a) MgO, (b) Mg$_2$B$_2$O$_5$ (monoclinic), (c) Kotoite Mg$_3$B$_2$O$_6$, and (d) Mg$_2$B$_2$O$_5$ (triclinic).  The Fermi energy level is represented by a vertical dashed red line.  The additional valence band peaks which form in the monoclinic and triclinic forms of Mg$_2$B$_2$O$_5$ are denoted by blue arrows.}  
\label{all_total_dos} 
\end{center} 
\end{figure}

Introducing boron into the MgO region will affect 
the density of states in the valence and conduction band.  In Figure \ref{all_total_dos}, the DOS of the three Mg-B-O systems is compared to MgO.  The MgO conduction band DOS gradually decreases as it approaches the conduction band edge.  In contrast, all three Mg-B-O systems show two sharp peaks at the conduction band edge.  In addition, the valence $s$ and $p$ band DOS in the monoclinic and triclinic forms of Mg$_2$B$_2$O$_5$ are split with the formation of two sharp secondary peaks at slightly lower energies denoted by arrows.  While the valence DOS for the $s$ and $p$ bands in Kotoite is slightly broadened in comparison to MgO, separate DOS peaks do not form.

\begin{figure}
\begin{center}
   \subfigure{
   \label{pdos_a}
   \includegraphics[angle=0,width=8cm]{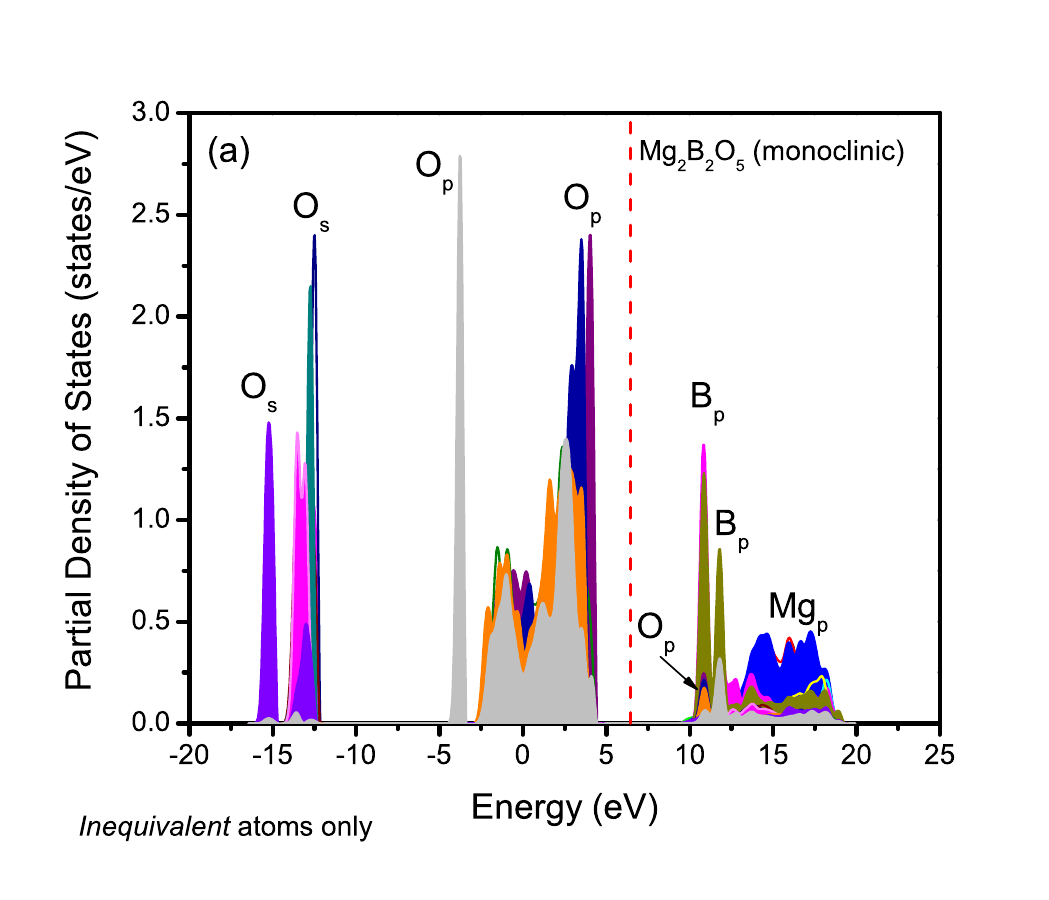} 
   } 
   \subfigure{
   \label{pdos_b}
   \includegraphics[angle=0,width=8cm]{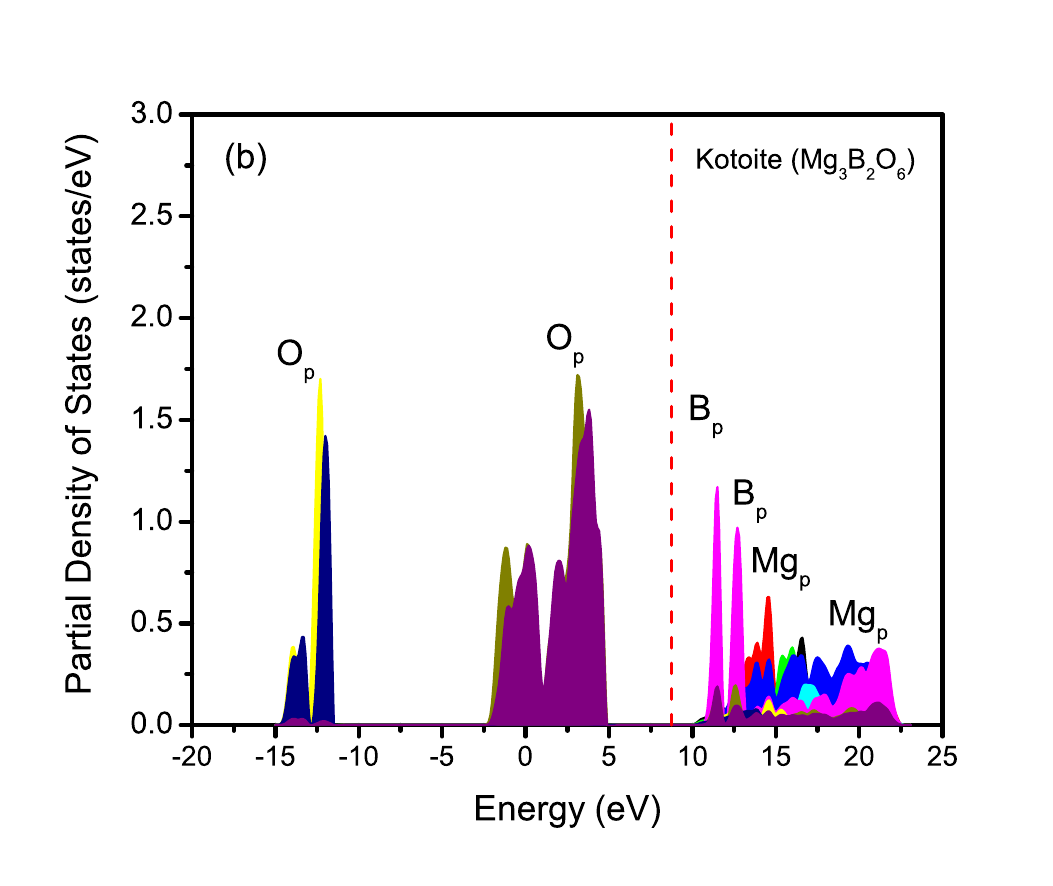}
   }
\caption{(color online) The partial density of states is shown for (a) monoclinic Mg$_2$B$_2$O$_5$ and (b) Kotoite Mg$_2$B$_3$O$_6$.  The partial density of states is only plotted for inequivalent atoms.  The Fermi energy level is denoted by a red dash line.}  
    \label{pdos_plot}
	\end{center}
\end{figure}

The cause of these DOS changes for MgB oxides can be determined by examining the partial DOS resolved into angular momentum channels (Fig. ~\ref{pdos_plot}).  The sharp peaks at the conduction band edge in all three Mg-B-O crystal structures are due to boron $p$ states.  In monoclinic and triclinic  Mg$_2$B$_2$O$_5$, the splitting of the $s$ and $p$ valence DOS is due to the formation of sharp peaks related to boron-oxygen bonding.  The monoclinic and triclinic forms of Mg$_2$B$_2$O$_5$ are characterized by double boron-oxygen triangles which are connected by an oxygen in the center.  This oxygen has a different bonding environment than the other oxygens that form the Mg octahedras.  This leads to splitting of the valence band DOS.  Kotoite, in contrast, only has a single BO$_3$ triangle and lacks this DOS splitting in the valence bands.

Although EELS data shows boron forming a relatively uniform distribution of BO$_3$ triangles in the MgO region\cite{cha_eels_apl_2007}, experimentally it is difficult to resolve whether this leads to Mg-B-O crystals or clusters of B$_2$O$_3$ in a MgO matrix.  To rule out B$_2$O$_3$ cluster formation, I also compared the total energy of the different Mg-B-O compounds to the formation of separate MgO and B$_2$O$_3$ regions.  The trigonal crystal structure for B$_2$O$_3$\cite{b2o3_crystal_structure} was used for total energy calculations.  The energy difference per atom in comparison to separate B$_2$O$_3$ and MgO regions is -0.0944 eV for Kotoite, -0.1005 eV for monoclinic Mg$_2$B$_2$O$_5$, and -0.1006 eV for triclinic Mg$_2$B$_2$O$_5$. In all cases, the Mg-B-O crystal structures are more stable than separate regions of MgO and B$_2$O$_3$.  This indicates the formation of isolated B$_2$O$_3$ clusters is unlikely and that the oxide region is made up of a crystalline magnesium boron oxide.  However, all three structures are fairly close in terms of total energy and other factors must be considered to determine the correct Mg-B-O structure.

\begin{figure}
\begin{center}
   \subfigure{
   \label{mgo_cband}
   \includegraphics[angle=0,width=8cm]{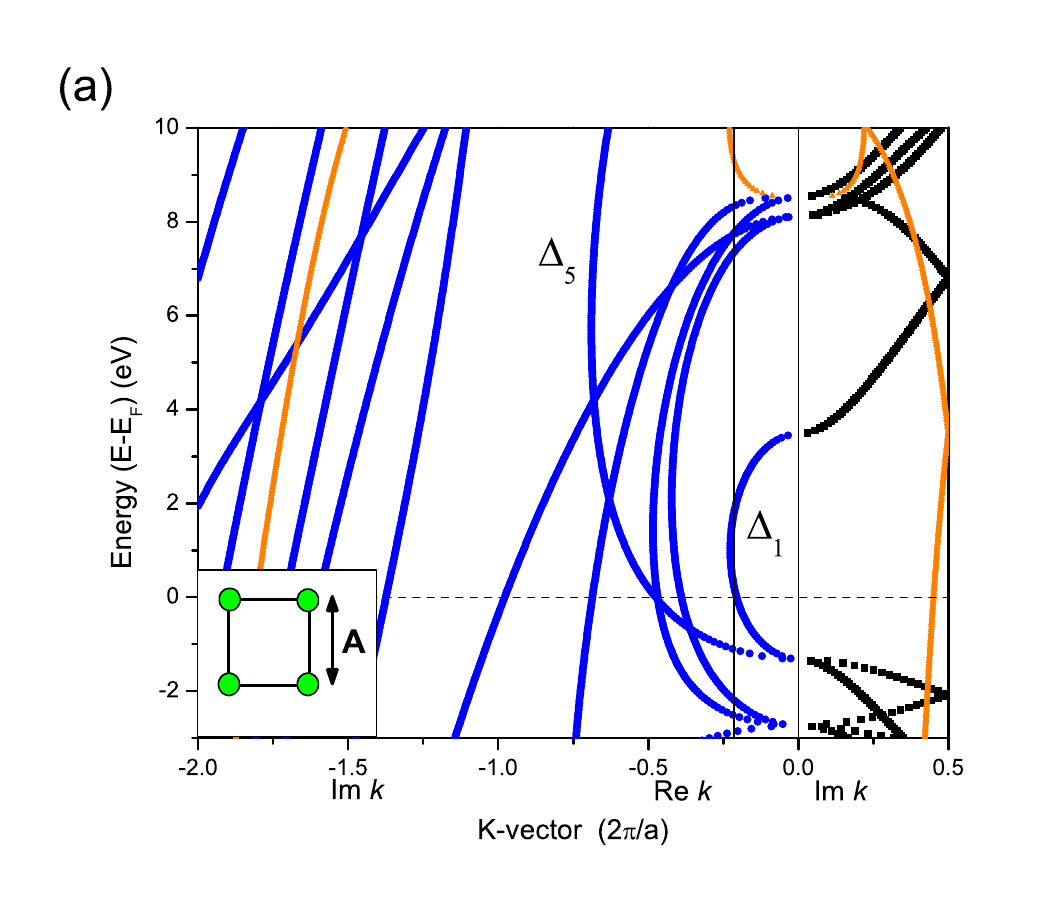}} 
   \subfigure{
   \label{kotoite_cband}
   \includegraphics[angle=0,width=8cm]{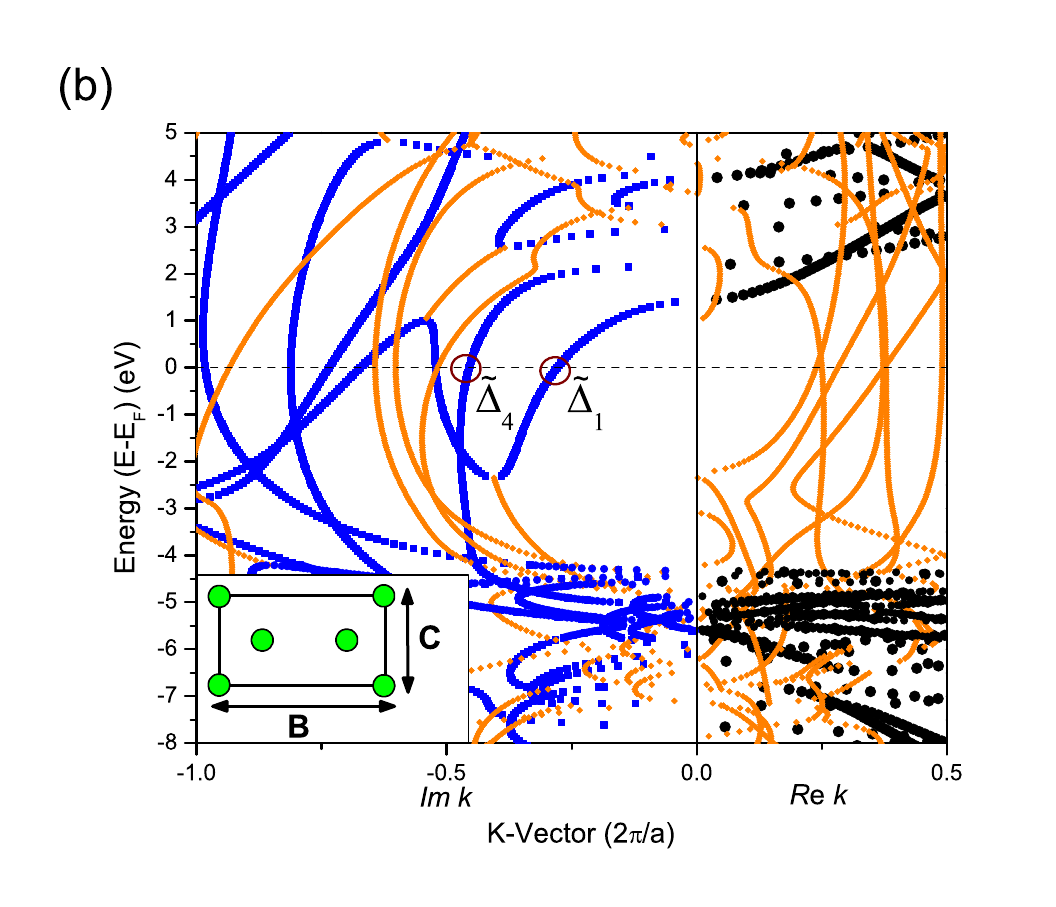}
   }
\caption{The complex band structure of MgO and Kotoite (Mg$_3$B$_2$O$_6$) are shown in panels a and b, respectively.  K-vector components are shown in terms of positive and negative regions of the plot to indicate real and imaginary components, respectively.  Real bands are denoted by black lines.  Purely imaginary bands are given by blue lines and complex bands are denoted by orange lines.  The (001) MgO surface and the (100) Kotoite surface are shown as insets in panels (a) and (b), respectively.  Mg atoms are denoted by green circles.}  
    \label{cband_plot}
	\end{center}
\end{figure}

Several key pieces of information support Kotoite formation in rf-sputtered MTJs.  The measured boron concentration in these Mg-B-O regions is fairly low (5-10\%)\cite{read_mgbo_barriers_apl_2009,cha_2009}.  This favors a Mg-B-O crystal with a low boron concentration.  Of the three systems considered in this work, Kotoite, with single BO$_3$ triangles, has the lowest boron concentration (18\%).  Experimental studies also indicate that the Mg-B-O region is crystalline after annealing and has a good epitaxial match with MgO on one side and still acts to template the neighboring FeBCo leads into bcc crystalline layers.  This indicates that the Mg-B-O crystal has a good lattice match with MgO in the (100) direction.  In Figure \ref{mgo_cband}, the MgO (100) crystal face is shown with Mg atoms in green.  The (100) face of Kotoite is shown in Figure \ref{kotoite_cband}.  Kotoite is orthorhombic and the (100) face is rectangular.  The lattice match with (001) MgO in this direction is very good.  The $B$ lattice vector of Kotoite is 8.416 $\AA$ which is very close to twice the MgO lattice constant ($2A_{MgO}=8.422 \AA$).  The $C$ lattice vector of Kotoite is 4.497 $\AA$ which is slightly larger than the MgO lattice vector, 4.211 $\AA$.  The Mg atomic positions on the (100) face of Kotoite also mimic those found in (001) MgO for almost all sites.  The monoclinic and triclinic forms of Mg$_2$B$_2$O$_5$ in contrast provide a poor lattice match to MgO and would be unable to template FeCo into bcc layers.

   
Zhang and Butler \textit{et al.} showed that the high TMR in FeCo/MgO/FeCo MTJs relies on symmetry selective filtering in MgO\cite{zhang_butler_prb_feco_2004}.  In particular, the FeCo majority carrier $\Delta_1$ band couples efficiently with the $\Delta_1$ complex band in MgO, while minority carriers can only couple through the $\Delta_5$ and $\Delta_2$ bands which decay much faster.  In order for Kotoite to provide high TMR values, it must provide similar symmetry selective filtering to that of MgO.  To determine whether this is the case, the complex band structures for both the (001) direction in MgO and the (100) direction in Kotoite were calculated using the approach of Choi and Ihm\cite{choi_and_ihm_complex_bands} as implemented in the Quantum Espresso package\cite{smogunov_complex_band_surface_science}.  The complex band structure for the tetragonal unit cell of (001) MgO is shown in Figure \ref{mgo_cband}.  In the (001) direction for MgO, the band symmetry group is $C_{4v}$.  As has been reported in previous studies, there is a substantial difference in the imaginary component of the $\Delta_1$ and $\Delta_5$ complex bands that leads to efficient majority spin filtering.  The complex band structure for Kotoite in the (100) direction is shown in Figure \ref{kotoite_cband}.  The band symmetry group along the $\Gamma$ to X line is $C_{2v}$.  Two different imaginary bands labeled $\tilde{\Delta}_1$ and $\tilde{\Delta}_4$ intersect the Fermi energy at -0.28 and -0.46 respectively.  These imaginary bands are continuations of real conduction bands based on boron $p$ states.  The imaginary component of the two bands determines how effectively electrons tunnel through the barrier.  The large separation in the imaginary components in these two bands indicates that electrons that can couple to the orthorhombic $\tilde{\Delta}_1$ band should dominate transport through the oxide region.  Introducing FeCo leads on either side of the oxide region will cause the position of the Fermi energy level to change.  However, over the entire energy range in the band gap, the imaginary components of $\tilde{\Delta}_1$ is significantly smaller than $\tilde{\Delta}_4$.



This interface presents an interesting problem in terms of symmetry filtering for MTJs.  The FeCo bcc layers and (001) MgO all possess cubic symmetry\cite{butler_prb_2001}.  However, Kotoite is orthorhombic and possesses C$_{2v}$ symmetry along the [100] direction.  This raises the question of how the cubic symmetry states ($\Delta_1$, $\Delta_5$, and $\Delta_2$) will couple into the $\tilde{\Delta}_1$ and $\tilde{\Delta}_4$ bands in this lower symmetry region.  The cubic C$_{4v}$ $\Delta_1$ bands transform as linear combinations of functions with $1$, $z$, and $2z^2-x^2-x^2$.  The C$_{2v}$ $\tilde{\Delta}_1$ bands in orthorhombic systems transform as linear combinations of $1$, $z$, $x^2$, $y^2$, and $z^2$.  Of the symmetry states available in Kotoite, the $\tilde{\Delta}_1$ band should couple most effectively with the majority spin $\Delta_1$ band in the FeCo magnetic leads.  For the minority carriers, the cubic C$_{4v}$ $\Delta_5$ band transforms as linear combinations of $zx$ and $zy$ symmetry.  The closest analogue in the orthorhombic system is the $\tilde{\Delta}_4$ band which transforms as $yz$ and has a much larger imaginary component than the orthorhombic $\tilde{\Delta}_1$ band.  Based on this symmetry analysis and the complex band structure, Kotoite should serve as an effective spin filter similar to MgO.


In this Letter, I have used density functional approaches to examine three possible magnesium boron oxides that could form during rf-sputtering of magnetic tunnel junctions.  Kotoite (Mg$_3$B$_2$O$_6$) is shown to be a strong candidate for symmetry based spin filtering.  Density of states analysis shows that all three compounds have sharp boron $p$ bands situated at the conduction band edge.
Several factors indicate that Kotoite is the best candidate for formation in MTJs.  The boron concentration in this structure is closest to that measured in experiments.  Kotoite grown in the (100) direction also provides a good epitaxial match to MgO (001) and can act to template neighboring FeCo regions as crystalline bcc layers.    The diffusion of boron into the MgO region may also help passivate Mg or O vacancies that would otherwise reduce the TMR in the magnetic tunnel junctions\cite{mgo_vacancies_mather_prb_2006, velev_apl_mgo_vacancies_2007}.  Complex band structure analysis shows that the cubic (001) $\Delta_1$ should couple with the orthorhombic (100) $\tilde{\Delta}_1$ bands in Kotoite, while the cubic (001) $\Delta_5$ bands should partially couple with the orthorhombic (100) $\tilde{\Delta}_4$ bands.  There is sufficient difference in the decay rates of the imaginary bands so that spin filtering should be expected.  This Letter has focused on the potential formation of Kotoite under present MTJ fabrication conditions.  Further experimentally research is required to determine conclusively whether Kotoite is indeed present in the barrier region.  Regardless of this issue, the current study shows that an experimental effort focused on forming MTJs based on Kotoite barriers is worthy of exploration.  Using a lower symmetry oxide region between two cubic symmetry magnetic leads could provide a new route for enhanced spin filtering and potentially a new generation of high TMR devices.   

\begin{acknowledgments}
Calculations were done on the Intel Cluster at the Cornell Nanoscale Facility, part of the National Nanotechnology Infrastructure Network (NNIN) funded by NSF.  Special thanks to John Read and Judy Cha for advice on MBO MTJ experimental issues, Alexander Smogunov for advice on complex band structures, and Volker Evert for access to his Augmented Spherical Wave code.  
\end{acknowledgments}






\end{document}